\documentclass[10pt, conference]{IEEEtran}
\IEEEoverridecommandlockouts
\usepackage{cite}
\usepackage{amsmath,amssymb,amsfonts}
\usepackage{graphicx}
\usepackage[dvipsnames]{xcolor}
\usepackage{booktabs}
\usepackage{pifont}
\usepackage[
  colorlinks=true,
  citecolor=blue,
  urlcolor=black,
  linkcolor=black
]{hyperref}

\definecolor{degcol}{HTML}{B02A2A}
\definecolor{spdcol}{HTML}{006300}
\newcommand{\degc}[1]{\textcolor{degcol}{#1}}
\newcommand{\spdc}[1]{\textcolor{spdcol}{#1}}

\newcommand{\nTsweepTau}{0.28}

\newcommand{\nTsweepwikitextCycSpan}{0.02}
\newcommand{\nTsweepwikitextDegSpan}{0.07}
\newcommand{\nTsweepwikitextTmax}{32\,768}
\newcommand{\nTsweepwikitextTmin}{2048}
\newcommand{\nTtokens}{8192}
\newcommand{\nTtwotokens}{16\,384}

\newcommand{\nandaCycWikiMax}{2.83}
\newcommand{\nandaCycWikiMin}{2.24}

\newcommand{\nandaHoursMax}{1.7}
\newcommand{\nandaHoursMin}{0.4}
\newcommand{\nandaItersMax}{25}
\newcommand{\nandaItersMin}{13}

\newcommand{\ncycFigna}{1.23}
\newcommand{\ncycFullMantissa}{1.45}

\newcommand{\nhwFpAreaPct}{55}

\newcommand{\nhwMxuAreaPct}{25.8}

\newcommand{\nhwPowInfPct}{3.5}
\newcommand{\nhwPowSel}{166.7}
\newcommand{\nhwPowSelPct}{50}

\newcommand{\nhwSysAreaPct}{3.7}

\newcommand{\nkamCycMainMax}{2.28}
\newcommand{\nkamCycMainMin}{2.25}
\newcommand{\nkamCycMax}{2.34}
\newcommand{\nkamCycMin}{2.11}

\newcommand{\nkamCycWikiMax}{2.29}
\newcommand{\nkamCycWikiMin}{2.12}

\newcommand{\nkamOracleShareMax}{97}
\newcommand{\nkamOracleShareMin}{80}

\newcommand{\nrobustCycles}{2.39}

\newcommand{\nsweepCycMax}{2.83}
\newcommand{\nsweepCycMin}{1.91}
\newcommand{\nsweepTauMax}{0.6}
\newcommand{\nsweepTauMin}{0.15}
\newcommand{\nsweepTauSafeMax}{0.32}
\newcommand{\ntaskAvgBase}{72.8}
\newcommand{\ntaskAvgKam}{72.9}
\newcommand{\ntaskDeltaMaxAbs}{0.1}

\newcommand{\ntaskN}{6}
\newcommand{\ntauFixed}{0.28}
\newcommand{\nvsqCycMax}{3.68}
\newcommand{\nvsqCycMin}{3.56}
\newcommand{\nvsqDegMax}{6.8}
\newcommand{\nvsqDegMin}{4.4}
\newcommand{\nxferAltCfourN}{6}
\newcommand{\nxferAltCfourOver}{1}

\newcommand{\nxferAltWikiOver}{5}

\newcommand{\nxferOffDiagonal}{42}
\newcommand{\nxferOverBudget}{18}
\newcommand{\nxferWorstDeg}{2.9}

\newcommand{\spelled}[1]{\ifcase#1 zero\or one\or two\or three\or four\or five\or six\or seven\or eight\or nine\else #1\fi}

\begin{document}

\title{Kameleo: In-Situ Activation-Precision Selection for Bit-Serial LLM Accelerators}

\author{
  \IEEEauthorblockN{Robin Geens\IEEEauthorrefmark{1}}
  \IEEEauthorblockA{MICAS, KU Leuven\\
  Leuven, Belgium}
  \and
    \IEEEauthorblockN{Thomas Debelle\IEEEauthorrefmark{1}}
    \IEEEauthorblockA{Independent researcher}
    \and
    \IEEEauthorblockN{Marian Verhelst}
    \IEEEauthorblockA{MICAS, KU Leuven\\
    Leuven, Belgium}
    \and
    \IEEEauthorblockN{Chao Fang}
    \IEEEauthorblockA{MICAS, KU Leuven\\
    Leuven, Belgium}
}

\maketitle
{\renewcommand{\thefootnote}{\fnsymbol{footnote}}%
\footnotetext[1]{Equal contribution.}}

\begin{abstract}
Weight-only quantization keeps the activations of large language models (LLMs) in a higher-precision format, pushing the on-chip inference cost to the activation arithmetic. Bit-serial, variable-length accelerators offer a promising path to reduce this cost by lowering activation precision while preserving model quality. They process only the mantissa bits that each layer needs. The length has to be chosen per layer and operator, and today it is chosen by offline calibration: a search driven by the perplexity of the full model on a representative dataset. The search is repeated for every model and workload, and its result degrades silently on other data. This paper presents Kameleo, an accelerator that selects the activation precision at runtime, on the accelerator, from the data it is processing. Kameleo replaces the global perplexity by a local proxy that compares the output of each GEMM at reduced and at full precision. It exploits the fact that a bit-serial datapath produces this comparison almost for free, because the reduced-precision result is an intermediate state of the full-precision computation. Two passes over the first tokens of a workload set the precision of every GEMM. On four LLMs and seven datasets, Kameleo stays within a 1\% perplexity budget at a \nkamCycMin--\nkamCycMax$\times$ reduction of prefill cycles on the matrix multiplication unit (MXU), close to the reduction of configurations calibrated offline on the evaluation data itself. Calibrated configurations transferred to other data, in contrast, often violate the budget. The selection hardware adds only \nhwSysAreaPct\% to the accelerator area. Kameleo thereby makes variable-length mantissas feasible in real deployments, where offline calibration for every model and workload is not.
\end{abstract}

\section{Introduction}
\label{sec:intro}

\begin{figure}[t]
    \centering
    \includegraphics[width=0.7\linewidth]{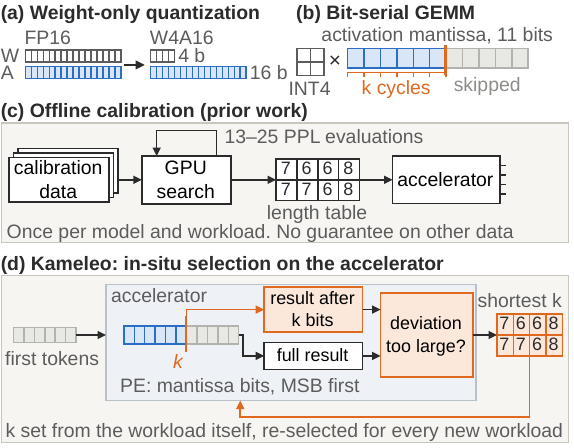}
    \caption{
    (a) Weight-only quantization. Activation arithmetic dominates.
    (b) Bit-serial GEMM: cutting the mantissa to $k$ bits saves cycles. $k$ must be chosen for each GEMM. 
    (c) Prior work chooses $k$ offline, once per model and workload. 
    (d) Kameleo selects $k$ in situ, on the accelerator. 
    }
    \label{fig:motivation}
\end{figure}

The predominant quantization strategy for today's large language models (LLMs) is weight-only quantization. Low-bit integer weights proportionally increase the throughput of the memory-bound decode stage, while keeping the activations in a higher-precision format (e.g., FP16) leaves the task accuracy intact~\cite{figna,frantarGPTQAccuratePostTraining2023,linAWQActivationawareWeight2024,shaoOmniQuantOmnidirectionallyCalibrated2024}. The higher-precision activation arithmetic is, however, far more complex than the integer weight arithmetic, so the on-chip bottleneck for energy and throughput shifts to the activations (Fig.~\ref{fig:motivation}a). Recent accelerators push the Pareto front further by pairing the low-precision weights with outlier-aware activation data types~\cite{anda,opal,olive,tender,m-ant}, and variable-length formats~\cite{anda,opal} give each general matrix multiplication (GEMM) only as many activation bits as it needs. On a bit-serial datapath the mantissa length translates directly into cycles, so every bit that a layer can do without is a proportional throughput gain in the compute-bound prefill phase (Fig.~\ref{fig:motivation}b).

The difficulty of a variable format is that its length has to be chosen, and existing work selects the precision through extensive offline calibration. The greedy search of ANDA~\cite{anda}, for instance, tries candidate lengths for every GEMM type and evaluates the perplexity (PPL) of the full model on a representative calibration dataset after every step. This regime has two shortcomings. First, calibration is compute-intensive, since every search step is a full-model evaluation, and ANDA needs \nandaItersMin{} to \nandaItersMax{} evaluations per model and dataset. Second, the result is data-dependent. The search has to be redone for every model and every workload, and the accuracy degrades silently when the deployment data drifts away from the calibration data, as Sec.~\ref{sec:shift} shows. Both make variable-length activation precision impractical in deployment (Fig.~\ref{fig:motivation}c).

This paper presents Kameleo, an LLM accelerator with variable-length activation precision that selects the precision at runtime, based on the data it processes (Fig.~\ref{fig:motivation}d). Online precision selection requires a quality signal that is local and label-free, in contrast to the global PPL used in offline calibration. Kameleo uses the deviation between the output of a GEMM at a reduced mantissa and the same output at full precision. We show that a bit-serial datapath produces this comparison almost for free, because the reduced-precision result is an intermediate state of the full-precision computation. Our contributions are as follows:
\begin{itemize}
    \item A local, label-free proxy for LLM quality under reduced activation precision, whose two parameters are fixed at design time and need no retuning per model or dataset (Sec.~\ref{sec:method}).
    \item A lightweight extension of a bit-serial processing element (PE) that captures the candidate results during the ordinary full-precision pass. Two passes over the first tokens of a workload then set every GEMM's precision in situ (Sec.~\ref{sec:hw}).
    \item An evaluation on four LLMs and seven datasets, in which Kameleo reaches \nkamOracleShareMin--\nkamOracleShareMax\% of the cycle reduction of configurations calibrated on the evaluation data itself and maintains model quality on data where offline calibrations fail (Sec.~\ref{sec:eval}).
\end{itemize}

\section{Runtime Precision Selection}
\label{sec:method}

\subsection{From a Global to a Local Quality Metric}
\label{sec:rule}

Offline calibration measures quality at the model output, as the PPL over a dataset. A precision setting is considered quality-neutral when it degrades the PPL by less than 1\%~\cite{anda,dettmersSpQRSparseQuantizedRepresentation2023,hooperFGMPFineGrainedMixedPrecision2025}. As a search signal, however, the PPL is poor: it needs labeled data, it is only available after the last layer, and one number for the whole model does not say which of the hundreds of GEMMs can reduce their precision. Runtime selection needs the opposite: a proxy local to one GEMM, so that all GEMMs can be set independently and in parallel. It should also be label-free, so that it can be evaluated on the deployment data itself, and cheap enough to compute next to the arithmetic. Kameleo's proxy is the deviation that a reduced mantissa causes in the output of the GEMM where it is applied. Let $y$ be an output element computed with the full $M$-bit mantissa and $\hat y_k$ the same element computed with the mantissa truncated to $k$ bits. The element counts as an \emph{error} when
\begin{equation}
    |\hat y_k - y| > 2^{-s}\,|y|,
    \label{eq:flag}
\end{equation}
and the \emph{error rate} $\varepsilon_k$ of the GEMM is the fraction of its outputs that are errors. A relative deviation below $2^{-s}$ means that $\hat y_k$ keeps the leading $s$ significant bits of $y$, whatever the number of input bits dropped. Counting errors instead of accumulating their magnitude keeps a few outliers from dominating and reduces the metric to one bit per output element. A length $k$ is \emph{accepted} when $\varepsilon_k \le \tau$.

\begin{figure}[t]
    \centering
    \includegraphics[width=\linewidth]{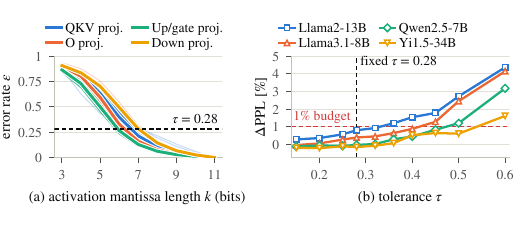}
    \caption{(a) Error rate against mantissa length for every GEMM of Llama2-7B on WikiText2 (thin: the 32 layers, thick: group mean). The mantissa length selected for a layer is where its curve crosses $\tau$. (b) PPL degradation against $\tau$ for the four models on WikiText2, with our fixed $\tau$ marked.}
    \label{fig:metric}
\end{figure}

The two parameters are properties of the acceptance rule itself, not of a model or a dataset. We set $s = 4$ and $\tau = \ntauFixed$ once and use them unchanged for every GEMM, model and dataset in this paper. This choice is further validated in Sec.~\ref{sec:eval-tau}. Fig.~\ref{fig:metric}a shows the metric at work: the error rate rises monotonically as the mantissa shortens, at a rate that differs per GEMM group and per layer. Every GEMM therefore obtains its own length where its curve crosses $\tau$. Fig.~\ref{fig:metric}b relates the local rule to the global budget: as $\tau$ grows, the PPL degradation of every model rises gradually. The fixed $\tau$ keeps all of them within the 1\% budget, and the budget is only crossed well beyond it. A single $\tau$ can hence be fixed at design time with margin.

\subsection{Search and Lock}
\label{sec:search}

Because the error rate is monotonic in the mantissa length (Fig.~\ref{fig:metric}a), the shortest accepted length of a GEMM is found by a search over the nine candidate lengths of 3 to 11 bits rather than an exhaustive sweep. All GEMMs are searched at once: every selection pass is an ordinary full-precision forward pass over $T$ tokens of the workload, in which each GEMM evaluates three candidate lengths. One pass narrows the nine candidates to at most three and a second pass resolves them, after which the lengths are locked and inference continues at reduced precision. Selection thus consumes $2T$ tokens of ordinary inference whose outputs are served as usual, and its trigger (per workload, periodically, or on a change of task) is a deployment policy.

\section{Hardware Architecture}
\label{sec:hw}

ANDA~\cite{anda} is a recent accelerator for weight-quantized LLMs. Its PEs multiply INT4 weights with block-floating-point (BFP) activations one mantissa bit-plane per cycle, with a mantissa length per GEMM type that is selected by offline calibration. We build Kameleo as a minimal extension of it: its bit-serial PE already produces, cycle by cycle, the partial sums that the proxy of Sec.~\ref{sec:method} needs. Keeping everything else identical isolates the cost of runtime selection in Sec.~\ref{sec:eval}.

\begin{figure}[t]
    \centering
    \includegraphics[width=\linewidth]{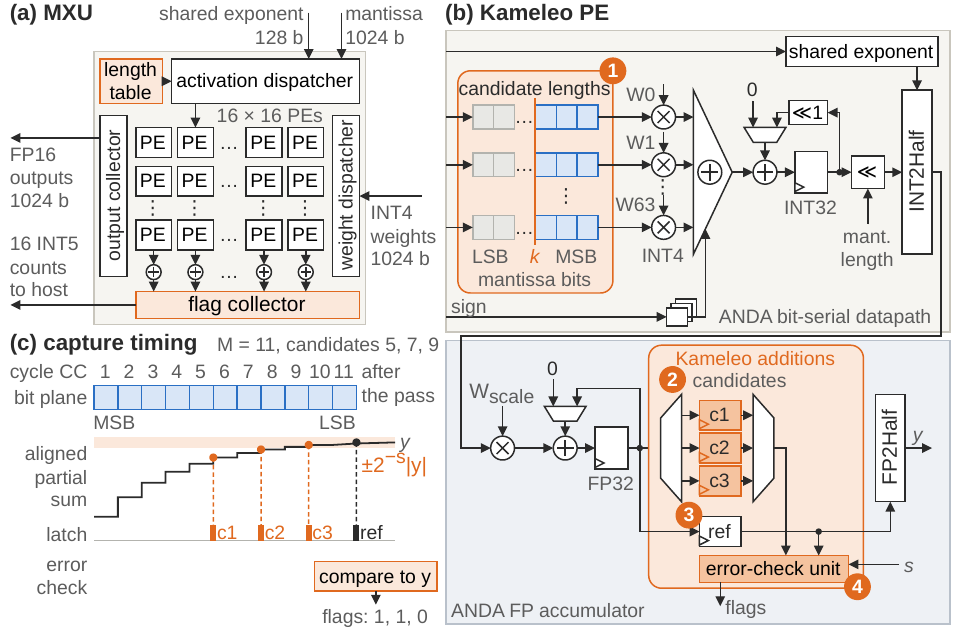}
    \caption{Kameleo hardware. (a) Matrix multiplication unit (MXU) with the added length table and flag collector. (b) PE: ANDA's bit-serial datapath and FP accumulator with the Kameleo additions \ding{202}--\ding{205}. (c) Capture timing for one output.}
    \label{fig:arch}
\end{figure}

\subsection{Processing Element}
\label{sec:pe}

The ANDA PE (Fig.~\ref{fig:arch}b) multiplies a group of 64 activation mantissas bit-plane by bit-plane with the INT4 weights, most-significant bit first. It sums the products in an adder tree, then shifts and accumulates the partial sum, so that after $k$ cycles it holds the exact dot product for a $k$-bit mantissa. The FP accumulator then scales this dot product with the shared exponent and the weight scale.

Obtaining the reduced and the full result of Eq.~\ref{eq:flag} nominally requires two passes, but the bit-serial order gives both in one (Fig.~\ref{fig:arch}c). After $k$ cycles, the partial sum shifted by the remaining $M-k$ bits is exactly the result for a $k$-bit mantissa. Three such results are captured in extra FP32 \emph{candidate registers} c1 to c3 (Fig.~\ref{fig:arch}b,c) when the bit counter reaches the candidate lengths of the current pass \ding{202}\ding{203}. The final result is captured in a \emph{reference register} \ding{204}. The alignment is folded into the exponent at latch time, and each candidate accumulates on the otherwise idle adder of the FP accumulator.

An \emph{error-check unit} \ding{205} evaluates Eq.~\ref{eq:flag} for every output element and each of c1 to c3 once the reference is available (Fig.~\ref{fig:arch}c). It forms $2^{-s}|y|$ by subtracting $s$ from the exponent of $y$, computes $|\hat y_k - y|$ with an FP32 subtractor and compares the two magnitudes, which yields one flag bit per candidate and element. Its registers are clock-gated outside the selection passes.

\subsection{Array and System Integration}
\label{sec:array}

The matrix multiplication unit (MXU) is a $16 \times 16$ array of these PEs (Fig.~\ref{fig:arch}a). A \emph{flag collector} sums the flags of the 16 PEs of a column into one INT5 count per candidate, so that the array emits 16 counts per candidate and output tile instead of 256 flag bits. The host controller accumulates the counts per GEMM into $\varepsilon_k$, compares them with $\tau$ and writes the next candidate set and finally the locked lengths to the activation dispatcher.

\section{Evaluation}
\label{sec:eval}

\subsection{Experimental Setup}
\label{sec:setup}

We evaluate Llama2-13B~\cite{touvronLlama2Open2023}, Llama3.1-8B~\cite{grattafioriLlama3Herd2024a}, Qwen2.5-7B~\cite{qwenQwen25TechnicalReport2025} and Yi1.5-34B~\cite{youngYiOpenFoundation2024} on WikiText2~\cite{merityPointerSentinelMixture2016}, C4~\cite{raffelExploringLimitsTransfer2023}, PG-19~\cite{raeCompressiveTransformersLongRange2019}, LAMBADA~\cite{papernoLAMBADADatasetWord2016}, CBT~\cite{hillGoldilocksPrincipleReading2016}, a corpus of Python source files (Code) and German Wikipedia (Wiki-DE), the last two deliberately far from English prose. The per-layer and per-dataset analyses (Fig.~\ref{fig:metric}a, Fig.~\ref{fig:transfer}) use Llama2-7B. The weights are quantized to INT4 using the quantizer of the OmniQuant framework~\cite{shaoOmniQuantOmnidirectionallyCalibrated2024} at its initialization, without learned clipping. We measure quality as the PPL over 2048-token windows, with a budget of 1\% PPL degradation relative to the W4A16 model with the full 11-bit BFP mantissa. We measure throughput as the reduction in MXU cycles of prefill workloads relative to W4A16 inference with FP16 activations. A bit-serial GEMM takes exactly its size times the activation mantissa length in cycles (16 for FP16 activations, 11 for the full BFP mantissa). This analytical count covers all GEMMs of the model, including the attention GEMMs, which keep the full mantissa, so the reported reductions hold for the complete model. We validated this count with a cycle-accurate simulation of the accelerator RTL.

ANDA selects one mantissa length per GEMM type (QKV, output, up/gate and down projection) with a greedy search that keeps the shortest lengths within the budget on a calibration dataset~\cite{anda}. We rerun this search for every model and dataset with the same weights and cost metric, at a cost of \nandaItersMin{} to \nandaItersMax{} full PPL evaluations, or \nandaHoursMin{} to \nandaHoursMax{} H100 GPU hours per search. Kameleo's selection is emulated in software with bit-exact mantissa truncation on the first $T = \nTtokens$ tokens of each dataset. The emulation uses a binary search, which selects the same lengths as the three-candidate hardware search under the monotonicity of Fig.~\ref{fig:metric}a. The PPL is then measured over the full dataset with the locked lengths. The hardware is written in SystemVerilog and synthesized with Cadence Genus on TSMC 16\,nm at 250\,MHz and 0.75\,V, like our ANDA implementation. Power is estimated from switching activity recorded in VCD files during simulation.

\subsection{Is the Tolerance a Calibration Parameter?}
\label{sec:eval-tau}

Fig.~\ref{fig:metric}b sweeps $\tau$ from \nsweepTauMin{} to \nsweepTauMax{} on WikiText2, which moves the MXU cycle reduction of the four models from \nsweepCycMin$\times$ to \nsweepCycMax$\times$. Their PPL degradation stays within the 1\% budget up to $\tau = \nsweepTauSafeMax$, so the fixed $\tau = \ntauFixed$ keeps a margin to the budget rather than being tuned to it. The selection prefix matters just as little: \nTsweepwikitextTmin{} instead of \nTsweepwikitextTmax{} tokens changes the cycle reduction by at most $\nTsweepwikitextCycSpan\times$ and the PPL by at most \nTsweepwikitextDegSpan\% (at $\tau = \nTsweepTau$). The tolerance thus behaves as a design constant, and Kameleo needs no calibration data of any kind.

\subsection{Quality and Throughput}
\label{sec:results}

\begin{table}[t]
    \centering
    \scriptsize
    \caption{WikiText2 PPL$\downarrow$ with \degc{degradation} relative to 11-bit BFP and \spdc{cycle reduction} relative to FP16 activations. $^\ddagger$BF16: overflows in FP16.}
    \label{tab:main}
    \setlength{\tabcolsep}{2pt}
    \renewcommand{\arraystretch}{1.5}
    \begin{tabular}{@{}lcccc@{}}
\toprule
Method & Llama2-13B & Llama3.1-8B & Qwen2.5-7B & Yi1.5-34B \\
\midrule
FP16 (no quantization) & $4.88$ & $6.24$ & $6.85$$^\ddagger$ & $4.90$ \\
W4A16, FP16 act. & $4.99^{\degc{-0.02\%}}_{\spdc{1.00\times}}$ & $6.75^{\degc{+0.02\%}}_{\spdc{1.00\times}}$ & $7.30^{\degc{+0.03\%}}_{\spdc{1.00\times}}$ & $5.01^{\degc{+0.01\%}}_{\spdc{1.00\times}}$ \\
FIGNA~\cite{figna}, 13-bit BFP & $4.99^{\degc{-0.02\%}}_{\spdc{1.23\times}}$ & $6.75^{\degc{+0.02\%}}_{\spdc{1.23\times}}$ & $7.30^{\degc{+0.02\%}}_{\spdc{1.23\times}}$ & $5.01^{\degc{+0.01\%}}_{\spdc{1.23\times}}$ \\
VS-Quant~\cite{daiVSQuantPervectorScaled2021}, INT4 act. & $5.21^{\degc{+4.41\%}}_{\spdc{3.61\times}}$ & $7.19^{\degc{+6.44\%}}_{\spdc{3.56\times}}$ & $7.79^{\degc{+6.85\%}}_{\spdc{3.62\times}}$ & $5.31^{\degc{+6.05\%}}_{\spdc{3.68\times}}$ \\
11-bit BFP (reference) & $4.99^{\degc{+0.00\%}}_{\spdc{1.45\times}}$ & $6.75^{\degc{+0.00\%}}_{\spdc{1.45\times}}$ & $7.29^{\degc{+0.00\%}}_{\spdc{1.45\times}}$ & $5.01^{\degc{+0.00\%}}_{\spdc{1.45\times}}$ \\
\midrule
ANDA~\cite{anda}, cal.\ WikiText2 & $5.04^{\degc{+1.00\%}}_{\spdc{2.37\times}}$ & $6.79^{\degc{+0.63\%}}_{\spdc{2.38\times}}$ & $7.36^{\degc{+0.95\%}}_{\spdc{2.24\times}}$ & $5.05^{\degc{+0.79\%}}_{\spdc{2.83\times}}$ \\
ANDA~\cite{anda}, cal.\ C4 & $5.11^{\degc{+2.36\%}}_{\spdc{2.57\times}}$ & $6.79^{\degc{+0.63\%}}_{\spdc{2.38\times}}$ & $7.36^{\degc{+0.95\%}}_{\spdc{2.24\times}}$ & $5.04^{\degc{+0.68\%}}_{\spdc{2.56\times}}$ \\
\textbf{Kameleo}, no cal.\ & $\mathbf{5.03}^{\degc{+0.80\%}}_{\spdc{2.29\times}}$ & $\mathbf{6.78}^{\degc{+0.39\%}}_{\spdc{2.29\times}}$ & $\mathbf{7.29}^{\degc{-0.04\%}}_{\spdc{2.12\times}}$ & $\mathbf{5.00}^{\degc{-0.16\%}}_{\spdc{2.26\times}}$ \\
\bottomrule
\end{tabular}

\end{table}

Table~\ref{tab:main} compares Kameleo on WikiText2 with fixed activation formats and with ANDA's calibrated configurations. The fixed formats mark the two ends of the trade-off. FIGNA~\cite{figna} keeps the accuracy of FP16 activations with a 13-bit mantissa, at a cycle reduction of only \ncycFigna$\times$. The per-vector INT4 activations of VS-Quant~\cite{daiVSQuantPervectorScaled2021}, re-evaluated in our flow, reach \nvsqCycMin{} to \nvsqCycMax$\times$ but degrade the PPL by \nvsqDegMin{} to \nvsqDegMax\%. The full 11-bit BFP mantissa of ANDA gives \ncycFullMantissa$\times$ at the accuracy of FP16 activations, and every further gain comes from shortening the mantissa per GEMM. ANDA calibrated on WikiText2 itself, a search that has seen the test distribution, reaches \nandaCycWikiMin{} to \nandaCycWikiMax$\times$ at a PPL degradation close to the 1\% budget. Kameleo reaches \nkamCycWikiMin{} to \nkamCycWikiMax$\times$ on all four models without any calibration and at a lower PPL degradation. The lower degradation follows from the local rule, which accepts a length only while most outputs of a GEMM keep their leading significant bits, a more conservative condition than the PPL budget. This margin also makes the selection robust under workload shift, as the next section shows. Downstream tasks confirm the quality result: over \spelled{\ntaskN} zero-shot benchmarks (PIQA, ARC-e, ARC-c, BoolQ, HellaSwag, WinoGrande~\cite{biskPIQAReasoningPhysical2020,clarkThinkYouHave,clarkBoolQExploringSurprising2019,zellersHellaSwagCanMachine2019,sakaguchiWinoGrandeAdversarialWinograd2019}) the average accuracy of every model moves by at most \ntaskDeltaMaxAbs{} points with Kameleo's WikiText2 lengths, for Llama2-13B from \ntaskAvgBase\% to \ntaskAvgKam\%.

\subsection{Benefit of Runtime Selection under Workload Shift}
\label{sec:shift}

\begin{figure}[t]
    \centering
    \includegraphics[width=\linewidth]{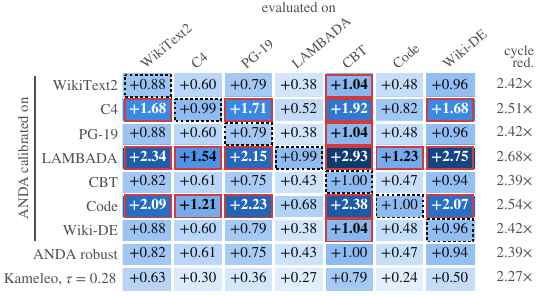}
    \caption{PPL degradation (\%) of Llama2-7B: ANDA calibrated on one dataset (dotted frames), the ANDA robust configuration built from all seven calibrations, and Kameleo. Red: over the 1\% budget.}
    \label{fig:transfer}
\end{figure}

To demonstrate that offline calibration on one dataset does not guarantee quality on others, we calibrate ANDA on each of the seven datasets in turn and evaluate every resulting configuration on all seven (Fig.~\ref{fig:transfer}). The dotted cells are the calibration datasets, and every other cell is a deployment on data that the calibration has not seen. Table~\ref{tab:main} already shows one such transfer, the C4 configuration of Llama2-13B failing on WikiText2, and Fig.~\ref{fig:transfer} shows that such failures are systematic. Every configuration meets the budget on its own data, but \nxferOverBudget{} of the \nxferOffDiagonal{} transfers exceed it, with PPL degradations of up to \nxferWorstDeg\%. The failures are not limited to the shifted workloads: the C4 configuration also fails on CBT, an English prose dataset like C4 itself. Which configuration transfers cannot be known in advance: on Yi1.5-34B the C4 configuration fails on only \spelled{\nxferAltCfourOver} of the \spelled{\nxferAltCfourN} other datasets, but the WikiText2 configuration fails on \spelled{\nxferAltWikiOver}.
A robust configuration, which selects the longest mantissa per GEMM type over all seven calibrations, restores the guarantee on those datasets at a cycle reduction of \nrobustCycles$\times$ over FP16 activations. It does so only because every one of those datasets was available for calibration in advance. Kameleo needs no such knowledge: selecting on each dataset itself with the same $\tau$, it meets the budget on every dataset at a cycle reduction of \nkamCycMainMin{} to \nkamCycMainMax$\times$. Offline calibration can therefore only guarantee the budget on the data it has seen, whereas runtime selection re-establishes it on whatever data the accelerator receives.

\subsection{Hardware Results}
\label{sec:hw-results}

\begin{figure}[t]
    \centering
    \includegraphics[width=\linewidth]{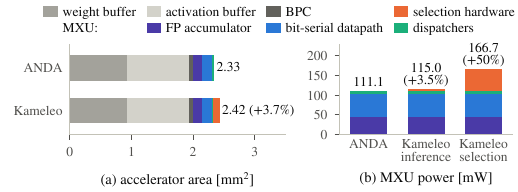}
    \caption{Synthesis on TSMC 16\,nm: (a) accelerator area, (b) MXU power for a full-precision GEMM.}
    \label{fig:hw}
\end{figure}

The candidate and reference registers and the error-check unit grow the area of the FP accumulator path by \nhwFpAreaPct\% and of the MXU by \nhwMxuAreaPct\% (Fig.~\ref{fig:hw}a). At the accelerator level, where the SRAM buffers dominate, the increase is \nhwSysAreaPct\%. Clock gating of the added registers outside the selection passes limits the MXU power increase during inference to \nhwPowInfPct\% (Fig.~\ref{fig:hw}b). During a selection pass the MXU power rises to \nhwPowSel\,mW, \nhwPowSelPct\% above ANDA, for the $2T = \nTtwotokens$ tokens of the two passes. Their outputs are still served, so a selection costs only the cycles not saved on them. Runtime selection is therefore almost free: a few percent of area and of inference power, and two ordinary prefill passes per selection, against \nandaItersMin{} to \nandaItersMax{} full-model evaluations per offline calibration.

\section{Conclusion}
\label{sec:conclusion}

Kameleo replaces the global PPL used for offline calibration by a local, label-free proxy read from the partial sums of a bit-serial PE. It selects the mantissa length of every GEMM in two passes on the accelerator. One fixed tolerance keeps four LLMs within a 1\% PPL budget on seven datasets, where transferred calibrations often fail, at \nkamCycMin--\nkamCycMax$\times$ fewer MXU cycles. The selection hardware adds \nhwSysAreaPct\% area and \nhwPowInfPct\% MXU power during inference. Kameleo thereby makes variable-length mantissas feasible in real deployments, where the workload is not known in advance and offline calibration for every model and workload is not.

\section*{Acknowledgment}
This project has been partly funded by the European Research Council (ERC) under grant agreement No.~101088865, the Flanders AI Research Program, Research Foundation Flanders (FWO) under grant No.~1S37125N, and long-term structural Methusalem funding by the Flemish Government.

\bibliographystyle{ieeetr}
\bibliography{ref}

\end{document}